\documentclass[%
superscriptaddress,
preprint,
 amsmath,amssymb,
 aps,
prd,
showkeys
]{revtex4-1}

\usepackage{amsmath, amsfonts, amssymb, mathrsfs, enumerate, graphicx, cases, hyperref}
\usepackage[T1]{fontenc}

\newcommand{\R}{{\mathbb R}}
\newcommand{\Z}{{\mathbb Z}}
\newcommand{\C}{{\mathbb C}}

\begin{document}


\title{Peeling property and asymptotic symmetries with a cosmological constant}

\author{Vee-Liem Saw}
\email{Vee-Liem@ntu.edu.sg}
\affiliation{Division of Physics and Applied Physics, School of Physical and Mathematical Sciences, 21 Nanyang Link, Nanyang Technological University, Singapore 637371}
\affiliation{Data Science and Artificial Intelligence Research Centre, Block N4 \#02a-32, Nanyang Avenue, Nanyang Technological University, Singapore 639798}

\author{Freeman Chee Siong Thun}
\noaffiliation

\begin{abstract}
This paper establishes two things in an asymptotically (anti-)de Sitter spacetime, by direct computations in the physical spacetime (i.e. with no involvement of spacetime compactification): (1) The peeling property of the Weyl spinor is guaranteed. In the case where there are Maxwell fields present, the peeling properties of both Weyl and Maxwell spinors similarly hold, if the leading order term of the spin coefficient $\rho$ when expanded as inverse powers of $r$ (where $r$ is the usual spherical radial coordinate, and $r\rightarrow\infty$ is null infinity, $\mathcal{I}$) has coefficient $-1$. (2) In the absence of gravitational radiation (a conformally flat $\mathcal{I}$), the group of asymptotic symmetries is trivial, with no room for supertranslations. 
\end{abstract}


\maketitle

\section{Introduction}

The peeling property of the (vacuum) curvature tensor for asymptotically flat spacetimes \cite{Sachs61,Bondi62,Sachs62} can be derived rather neatly using the Newman-Penrose formalism. This was done by employing a special null tetrad adapted to outgoing null cones and by supposing that the dyad component of the Weyl spinor $\Psi_0$ has a fall-off of $O(r^{-5})$ (together with some technicalities on differentiabilities) \cite{newpen62}. Using the Newman-Penrose equations and this single assumption on $\Psi_0$, one can then deduce the fall-offs for the spin coefficients, the unknown functions defining the null tetrad as well as the remaining dyad components of the Weyl spinor. In particular, a subset of the Bianchi identities with the $D\Psi_j$ derivatives ($j=1,2,3,4$), i.e. derivatives along the outgoing null cones, possess a nice hierarchical structure leading to $\Psi_n=O(r^{n-5})$ for the dyad components of the Weyl spinor, where $n=0,1,2,3,4$. A follow-up work by Newman and Unti solved for the behaviour of empty asymptotically flat spacetimes \cite{newunti62}.

Recently, the behaviour of empty asymptotically (anti-)de Sitter spacetimes has been studied \cite{Vee2016}, and further extended to contain Maxwell fields \cite{Vee2017} (see also \cite{Vee2017b,Vee2017d,Vee2018} \footnote{In the latest development, we found that gravitational waves radiated by a \emph{quadrupole source} must carry a \emph{positive-definite energy} in our Universe with $\Lambda>0$, within the \emph{full non-linear} theory of general relativity \cite{Vee2018}.}). These took the approach of exclusively studying the physical spacetime, i.e. no conformal rescaling was explicitly made in deriving those asymptotic solutions. In accordance to the case for asymptotically flat spacetimes, the imposed condition $\Psi_0=O(r^{-5})$ leads to the expected peeling property for the Weyl spinor and a similar condition on the dyad component of the Maxwell spinor $\phi_0=O(r^{-3})$ yields the analogous peeling behaviour for the Maxwell fields. These peeling properties of $\Psi_n$ and $\phi_m$ (where $m=0,1,2$) with a non-zero cosmological constant should not be surprising, as the peeling property for the Weyl spinor has been proven for (weakly) asymptotically simple spacetimes that do include asymptotically (anti-)de Sitter spacetimes \cite{Pen63,Pen65,Pen88}. This follows from admitting a smooth conformal compactifiability and working with a conformally compactified version of the spacetime, involving spinors.

The driving motivation in Ref. \cite{Vee2016} was to generalise the notion of the Bondi mass of an isolated gravitating system, and the mass loss due to energy carried away by gravitational waves with a positive cosmological constant, $\Lambda>0$ --- since we now know that our universe expands at an accelerated rate (which is well described by Einstein's theory with $\Lambda>0$). Although Bondi et al. led the initial breakthrough to this problem for the asymptotically flat case \cite{Bondi60,Bondi62}, the mass-loss formula for $\Lambda=0$ could also be obtained using the Newman-Penrose formalism (as shown by Newman and Unti \cite{newunti62}). Therefore, Ref. \cite{Vee2016} aimed to derive the asymptotic solutions to the Newman-Penrose equations with $\Lambda$ \`a la Newman and Unti. The mass-loss formula then arises from the Bianchi identity involving the $D'$ derivative of $\Psi_2$ (where $\Psi_2$ is the dyad component of the Weyl spinor with boost weight 0).


A key guidance that played a helpful role (towards the successful derivation of the asymptotic solutions with $\Lambda$) in Ref. \cite{Vee2016} was to closely check against the known asymptotic solutions for the asymptotically flat case, i.e. to work with coordinates and a null tetrad that would readily reduce to the well-known asymptotic solutions summarised at the end of Section 9.8 in Ref. \cite{Pen88} --- when $\Lambda$ is set to zero. To do so, one would need a coordinate system that is valid for \emph{any} $\Lambda\in\R$. It was noted that it is advantageous to first study the Schwarzschild-(anti-)de Sitter spacetime expressed in \emph{spherical coordinates}, since this coordinate system holds for any $\Lambda\in\R$ \cite{GP}. A coordinate transformation then introduced the desired retarded null coordinate $u$ in place of the time coordinate $t$, and thus the radial coordinate $r$ would approach null infinity $\mathcal{I}$ as $r\rightarrow\infty$. Subsequently, a Newman-Unti-like null tetrad (most notably such that $\tilde{l}=\tilde{d}u$, giving $\vec{l}=\vec{\partial}_r$) could be defined, and hence the spin coefficients were computed. This provided information on what the fall-offs for the spin coefficients would probably look like in general --- in particular, $\rho'$ and $\gamma$ both have terms of order $O(r)$ due to a non-zero $\Lambda$. Besides that, the function $U$ that appears in the null tetrad vector $\vec{n}$ is now $O(r^2)$ instead of $O(1)$ (also due to a non-zero $\Lambda$). Furthermore, this static spacetime indicated at which orders the mass term would begin to crop up in the various quantities.

With this, one could attempt to work out the asymptotic solutions using such fall-offs for the spin coefficients and the unknown functions in the null tetrad. It was soon discovered however (whilst working through the 38 Newman-Penrose equations) that the leading order term of the complex shear $\sigma$ (which encodes the physics representing the Bondi news for gravitational waves) would be zero or has to satisfy some unusual constraint equation --- unless these two spin coefficients $\sigma',\kappa'$ have the fall-offs: $\sigma'=O(1)$ and $\kappa'=O(1)$ (due to a non-zero $\Lambda$ \emph{and} a non-zero Bondi news) \footnote{These two spin coefficients $\sigma'$ and $\kappa'$ necessarily vanish for the Schwarzschild-(anti-)de Sitter spacetime (which is spherically symmetric) since they have non-zero spin weights. Well, the static case only served as an investigation that suggested what a suitable set of fall-offs might be. The eventual validation for the correct set of fall-offs came from being able to successfully generalise the asymptotic solutions for $\Lambda=0$ \cite{newunti62,Pen88} to include a non-zero $\Lambda$ with the physics of gravitational waves \cite{Vee2016}. This incidentally, does not require polyhomogeneous terms.}. The conclusion in Ref. \cite{Vee2016} was, one could successfully obtain the general asymptotic solutions for asymptotically (anti-)de Sitter spacetimes by stipulating that all the unknown functions could be expanded as inverse powers of $r$ away from $\mathcal{I}$ with sufficiently many orders, using the given fall-offs of the spin coefficients and the null tetrad (thus producing the mass-loss formula from the relevant Bianchi identity, with a proposal for the Bondi mass that includes $\Lambda$). On top of that, it was assumed that $\Psi_0=O(r^{-5})$, in accordance to the situation for asymptotically flat spacetimes. Whilst all these might have seemed like an ad hoc ansatz, they allowed one to sidestep the mathematical technicalities in \emph{deriving} the fall-offs (which in our opinion, was not the most pressing question with regards to discovering the \emph{physical results} of the mass-loss formula with $\Lambda>0$). Therefore, the setup assumed in Ref. \cite{Vee2016} may not be minimal.

It turns out that around the time this work was carried out, Szabados and Tod had just studied and presented the conformal behaviour of de Sitter-like spacetimes, working out the fall-offs for the physical metric and the spin coefficients (albeit in a different null tetrad, corresponding to a symmetric scaling over the spinor dyads $o^A$ and $\iota^A$) \cite{Szabados}. The stipulations made in Ref. \cite{Vee2016} may be checked and found to be consistent with those results given in Ref. \cite{Szabados} \footnote{The $\vec{l}$ and $\vec{n}$ null tetrad vectors in Ref. \cite{Szabados} are related to those in Refs. \cite{Vee2016,Vee2017} by a boost transformation. Apart from that, the $\vec{m}$ and $\vec{\bar{m}}$ null tetrad vectors in the former are prescribed to have no $\vec{\partial}_r$ component, whilst those in the latter papers are instead required to be parallel transported along $\vec{l}$ --- in the spirit of Refs. \cite{newpen62,newunti62}. These can be related via a null rotation around $\vec{l}$. Therefore, some work is required to convert the fall-offs from one set of null tetrad to the other. Note also that a smooth conformal compactifiability would rule out polyhomogeneous terms --- and is adequate to describe gravitational radiation.}. Whilst the results in Ref. \cite{Szabados} are focussed on $\Lambda>0$, those asymptotic solutions in Refs. \cite{Vee2016,Vee2017} hold for any $\Lambda\in\R$.

The main purpose of this paper is to show that the assumption on $\Psi_0$ being $O(r^{-5})$ made in Ref. \cite{Vee2016} is actually unnecessary. Instead, this is a consequence of a non-zero cosmological constant in the $D'\Psi_0$ Bianchi identity, which gives rise to an equation at order $r^{-3}$: $\Lambda(\Psi_0)^o_4/6=0$ [see Eq. (\ref{Psi4}) below]. Ergo, a non-zero cosmological constant together with the fall-off specifications of the unknown functions in the null tetrad and the spin coefficients, as well as the Weyl spinor vanishing on $\mathcal{I}$, \emph{will imply that $\Psi_0$ must have a fall-off of $O(r^{-5})$} --- so the peeling property follows, as worked out in Ref. \cite{Vee2016}. If the cosmological constant is zero however, one cannot deduce that $\Psi_0$ must be $O(r^{-5})$ according to these arguments since Eq. (\ref{Psi4}) is trivially satisfied. The corresponding peeling property for the Maxwell fields similarly follows \cite{Vee2017} \footnote{We note that with a cosmological constant, the peeling property is not quite an invariant concept, since null infinity $\mathcal{I}$ is not a null hypersurface for $\Lambda\neq0$ and so there is no well-defined null direction \cite{Pen88}. Nevertheless within the context of an isolated system with $\Lambda>0$, this isolated system defines a timelike infinity $i^+$ on the spacelike $\mathcal{I}$ and has a null hypersurface defining its cosmological horizon. The peeling property is then well-defined, with respect to this isolated system and the radiation emitted by it. See Ref. \cite{Vee2017d} for a discussion.}. The setup to this is given in the next section, with the proof of the peeling property given in Section 3. This result serves to show that the approach purely within the physical spacetime is consistent with that of the conformal approach in terms of the stipulated fall-offs when $\Lambda\neq0$.

A related asymptotic property of spacetime is its asymptotic symmetries. By directly calculating the asymptotic coordinate transformations that preserve the asymptotic fall-offs, we find that in the case with no gravitational radiation given off by an isolated system (this is the situation where $\mathcal{I}$ is conformally flat, i.e. its Bach tensor is zero), then the asymptotic group is trivial. The key motivation for doing these calculations is to eventually find out what the asymptotic group of coordinate transformations is, in a general spacetime \emph{with} gravitational radiation. However in that case, the equations become too intractable to be analytically solvable, as we have attempted. To solve it for the no-gravitational radiation case itself is a pretty challenging but fortunately achievable task, as we present in this paper --- with the specification of axisymmetry. This problem is of immense significance, because the asymptotic symmetry group forms the basis of a Hamiltonian formulation for the mass of an isolated system, in an asymptotically flat universe \cite{200years}. We work through the computations of the coordinate transformations which preserve the form of the asymptotic fall-offs with a cosmological constant in Section 4, which are directly analogous to the computations done by Bondi et al. in Ref. \cite{Bondi62}. As we specialise to the case with no-gravitational radiation in the following section, the equations to be solved for the coordinate transformations are a pair of partial differential equations (PDE) Eqs.\ (\ref{firstonetosolve})-(\ref{otheronetosolve}). One is the Laplace equation, with the other being a strange looking PDE. The method to solve these PDE that we come up with is non-trivial (perhaps the trickiest part involves solving a \emph{system} of linear inhomogeneous ordinary differential equations simultaneously, with \emph{unspecified} inhomogeneous functions), but it miraculously turns out to be solvable with only trivial solutions.

The problem of asymptotic symmetries has recently gained interest, especially with the inclusion of $\Lambda$. Refs. \cite{Mao19,Com19} studied the asymptotic structure with some relaxation of the fall-offs compared to those presented in Refs.\ \cite{Vee2016,Szabados}. For instance, Ref.\ \cite{Mao19} allowed for the inclusion of Robinson-Trautman waves by letting the $U^o_{-1}$ term be non-zero. This extension is beyond the consideration of Refs.\ \cite{Vee2016,Szabados} which focussed on the gravitational radiation given off by a compact isolated system. In particular with this physical setup, the $U^0_{-1}$ term can be eliminated by a gauge freedom in $(\nabla_a\Omega)(\nabla^a\Omega)=\Lambda/3+O(\Omega^2)$ [see section 4.2 in Ref.\ \cite{Szabados} that discussed this and the resulting metric in Eq.\ (4.18) of that paper]. Nevertheless, Ref.\ \cite{Mao19} found that even with their relaxed fall-offs with non-zero cosmological constant, there are no supertranslations --- consistent with the results presented here. Interestingly, Ref.\ \cite{Com19} arrived at new boundary conditions, with correction terms to the Bondi mass when $\Lambda\neq0$. However, we would like to emphasise the importance of the physical setup in Ref.\ \cite{Vee2016} where no source in the stress-energy tensor and no stray gravitational waves are permitted, other than those arising from the compact isolated source. In this physically important setup, we must be able to quantify the Bondi mass and the mass-loss formula due to the outgoing gravitational waves \emph{exclusively arising from this isolated source}. Ref.\ \cite{Vee2018} in fact established positivity of the energy carried by these outgoing waves arising from quadrupole sources in our universe with $\Lambda>0$ within the full non-linear general relativity, and proposed conditions to ensure positivity for general sources --- without requiring any correction to the Bondi mass from the asymptotically flat case.

This paper concludes with some comments and discussion on recent related work in the literature on gravitational radiation, putting them into perspective with regards to the peeling property with a cosmological constant.

\section{The setup}

Asymptotically (anti-)de Sitter spacetimes may be described by the following null tetrad \cite{Vee2016,Vee2017}
\begin{eqnarray}
\vec{l}&=&\vec{\partial}_r\\
\vec{n}&=&\vec{\partial}_u+U\vec{\partial}_r+X^\mu\vec{\partial}_\mu\\
\vec{m}&=&\omega\vec{\partial}_r+\xi^\mu\vec{\partial}_\mu\\
\vec{\bar{m}}&=&\bar{\omega}\vec{\partial}_r+\overline{\xi^\mu}\vec{\partial}_\mu,
\end{eqnarray}
where
\begin{eqnarray}
U&=&U^o_{-2}r^2+O(r)\label{U}\\
X^\mu&=&O(r^{-1})\label{X}\\
\omega&=&O(r^{-1})\\
\xi^\mu&=&O(r^{-1})\label{xi}.
\end{eqnarray}
Here, the Greek index $\mu$ denotes the two general angular coordinates $\theta, \phi$, which label the null geodesic generators of the outgoing null cones defined by $u=$ constant.

The directional derivatives are
\begin{eqnarray}
D&=&l^a\nabla_a=\frac{\partial}{\partial r}\\
D'&=&n^a\nabla_a=\frac{\partial}{\partial u}+U\frac{\partial}{\partial r}+X^\mu\frac{\partial}{\partial x^\mu}\\
\delta&=&m^a\nabla_a=\omega\frac{\partial}{\partial r}+\xi^\mu\frac{\partial}{\partial x^\mu}\\
\delta'&=&\bar{m}^a\nabla_a=\bar{\omega}\frac{\partial}{\partial r}+\overline{\xi^\mu}\frac{\partial}{\partial x^\mu}.
\end{eqnarray}

The spin coefficients are \footnote{The different fall-offs appearing here, as compared to the asymptotically flat case, are $\rho'^o_{-1}$, $\gamma^o_{-1}$, as well as the $O(1)$ order terms for $\sigma'$ and $\kappa'$. The first two are expected from the case of purely vacuum (anti-)de Sitter spacetime, where $\rho'^o_{-1}=-\Lambda/6$ and $\gamma^o_{-1}=-\Lambda/6$. The requirement of an $O(1)$ term for $\sigma'$ was found from the spin coefficient equation involving $D\sigma'$, otherwise the leading order term for the shear $\sigma$ that represents outgoing gravitational waves would vanish. After a substantially long sequence of calculations, it was found that a strange condition on the leading order term for $\sigma$ came up, which went away if $\kappa'$ has an $O(1)$ term. Again, those new leading order terms due to $\Lambda$ are consistent with the results presented in Ref. \cite{Szabados}.}
\begin{eqnarray}
\kappa&=&\gamma'=\tau'=0,\label{spincoe1}\\
\rho&=&\rho^o_1r^{-1}+O(r^{-3}), \rho^o_1\neq0\\
\rho'&=&\rho'^o_{-1}r+O(1), \rho'^o_{-1}\neq0\\
\sigma&=&\sigma^o_1r^{-1}+O(r^{-2})\\
\sigma'&=&O(1)\\
\gamma&=&\gamma^o_{-1}r+O(1), \gamma^o_{-1}\neq0\\
\alpha&=&O(r^{-1})\\
\alpha'&=&O(r^{-1})\\
\tau&=&O(r^{-1})\\
\kappa'&=&O(1).\label{spincoe2}
\end{eqnarray}
Here, $\kappa=0$ since $\vec{l}$ satisfies $l^a_{\ ;b}l^b=0$, whereas $\gamma'=\tau'=0$ by the condition that $\vec{n}, \vec{m}, \vec{\bar{m}}$ are parallel transported along $\vec{l}$. The $r^{-2}$ order of $\rho$ can be made zero by appropriately choosing the origin of $r$, which in this case is also an affine parameter of the congruence of null geodesics where $\vec{l}$ is tangent to. Each of these leading order terms $\rho^o_1$, $\rho'^o_{-1}$, $\gamma^o_{-1}$ cannot be zero, as the purely vacuum (anti-)de Sitter spacetime has such non-zero terms.

For the Weyl and Maxwell spinors, they must vanish at infinity [otherwise spacetime is not asymptotically (anti-)de Sitter] \footnote{Incidentally, a smooth conformal compactifiability and being vacuum (possibly with $\Lambda$) near $\mathcal{I}$ would imply the vanishing of the Weyl curvature on $\mathcal{I}$ \cite{Pen88}.}. So in general, they would have the following asymptotic expansions
\begin{eqnarray}
\Psi_0&=&(\Psi_0)^o_1r^{-1}+(\Psi_0)^o_2r^{-2}+(\Psi_0)^o_3r^{-3}+(\Psi_0)^o_4r^{-4}+O(r^{-5})\label{Weyl1}\\
\Psi_1&=&(\Psi_1)^o_1r^{-1}+(\Psi_1)^o_2r^{-2}+O(r^{-3})\\
\Psi_2&=&(\Psi_2)^o_1r^{-1}+O(r^{-2})\\
\Psi_3&=&O(r^{-1})\\
\Psi_4&=&O(r^{-1})\label{Weyl2}\\
\phi_0&=&(\phi_0)^o_1r^{-1}+(\phi_0)^o_2r^{-2}+O(r^{-3})\\
\phi_1&=&O(r^{-1})\\
\phi_2&=&O(r^{-1}).
\end{eqnarray}

With the Maxwell fields involved, we need to impose the condition $\rho^o_1=-1$ which ensures the usual behaviour of the asymptotic divergence of the congruence of null geodesics induced by $\vec{l}$ near $\mathcal{I}$. Otherwise, a non-zero $(\phi_0)^o_1$ would affect the divergence behaviour of these null geodesics near $\mathcal{I}$ (see the remark at the end of the next section). Then, we show in the next section that we have the following result:
\begin{gather}
\sigma^o_1=0, (\phi_0)^o_1=0, (\Psi_0)^o_1=0, (\Psi_0)^o_2=0, (\Psi_0)^o_3=0, \rho'^o_{-1}=-\frac{\Lambda}{6}, (\Psi_1)^o_1=0, (\Psi_1)^o_2=0,\nonumber\\
\gamma^o_{-1}=-\frac{\Lambda}{6}, (\Psi_2)^o_1=0, U^o_{-2}=\frac{\Lambda}{6},
\end{gather}
using the spin coefficient equations and the metric equations, as described below by solving for the relevant orders of $r$. Subsequently, one of the Maxwell equations can be used to show that $(\phi_0)^o_2=0$, which would imply the peeling property for the Maxwell spinor \cite{Vee2017}. Similarly, one of the Bianchi identities can then be used to show that $(\Psi_0)^o_4=0$, upon which the peeling property for the Weyl spinor follows \cite{Vee2016,Vee2017}.

Incidentally, the spacetime has inverse metric components:
\begin{align}
g^{ab}&=
\begin{pmatrix}
  0 & 1 & 0 & 0 \\
  1 & 2(U-|\omega|^2) & X^\theta-2\textrm{Re}(\xi^\theta\bar{\omega}) & X^\phi-2\textrm{Re}(\xi^\phi\bar{\omega}) \\
  0 & X^\theta-2\textrm{Re}(\xi^\theta\bar{\omega}) & -2|\xi^\theta|^2 & -2\text{Re}(\xi^\theta\overline{\xi^\phi}) \\
	0 & X^\phi-2\textrm{Re}(\xi^\phi\bar{\omega}) & -2\text{Re}(\xi^\theta\overline{\xi^\phi}) & -2|\xi^\phi|^2
\end{pmatrix}\\
&=
\begin{pmatrix}
  0 & 1 & 0 & 0 \\
  1 & \frac{\Lambda}{3}r^2+O(1) & O(r^{-1}) & O(r^{-1}) \\
  0 & O(r^{-1}) & O(r^{-2}) & O(r^{-2}) \\
	0 & O(r^{-1}) & O(r^{-2}) & O(r^{-2})
\end{pmatrix}.
\end{align}
For the special case of the Schwarzschild-(anti-)de Sitter spacetime, this becomes
\begin{align}
g^{ab}&=
\begin{pmatrix}
  0 & 1 & 0 & 0 \\
  1 & \frac{\Lambda}{3}r^2-1+\frac{2M}{r} & 0 & 0 \\
  0 & 0 & -\frac{1}{r^2} & 0 \\
	0 & 0 & 0 & -\frac{1}{r^2}\csc^2{\theta}
\end{pmatrix},
\end{align}
and the metric is
\begin{align}
g=-\left(\frac{\Lambda}{3}r^2-1+\frac{2M}{r}\right)du^2+2dudr-r^2(d\theta^2+\sin^2{\theta}d\phi^2).
\end{align}

\section{Proof of the peeling properties for the Weyl and Maxwell spinors}

In this section, we solve the Newman-Penrose-Maxwell equations order by order, like how it was done in Refs.\ \cite{Vee2016,Vee2017}. Listed below are the steps in the sequence of the relevant equations that are being solved. For each of these equations, the solutions from the lowest order up to the relevant orders are stated. The full list of equations and how they are completely solved are found in Refs.\ \cite{Vee2016,Vee2017}.

$D\sigma'=\sigma'\rho+\rho'\bar{\sigma}-k\phi_2\bar{\phi}_0$,
\begin{eqnarray}
1&:&\sigma^o_1=0,\textrm{ since }\rho'^o_{-1}\neq0\textrm{ when }\Lambda\neq0.\label{32}
\end{eqnarray}

$D\rho=\rho^2+\sigma\bar{\sigma}+k\phi_0\bar{\phi}_0$,
\begin{eqnarray}
r^{-2}:(\phi_0)^o_1=0.\label{Drho}
\end{eqnarray}

$D\sigma=2\rho\sigma+\Psi_0$,
\begin{eqnarray}
r^{-1}&:&(\Psi_0)^o_1=0\\
r^{-2}&:&(\Psi_0)^o_2=0\\
r^{-3}&:&(\Psi_0)^o_3=0.
\end{eqnarray}

$\displaystyle D\rho'=\rho'\rho+\sigma'\sigma-\Psi_2-\frac{\Lambda}{3}$,
\begin{eqnarray}
1:\rho'^o_{-1}=-\frac{\Lambda}{6}.
\end{eqnarray}

$D\alpha'=\alpha'\rho-\alpha\sigma-\Psi_1$,
\begin{eqnarray}
r^{-1}&:&(\Psi_1)^o_1=0\\
r^{-2}&:&(\Psi_1)^o_2=0.
\end{eqnarray}

$\displaystyle D\gamma=\tau\alpha-\bar{\tau}\alpha'+\Psi_2-\frac{\Lambda}{6}+k\phi_1\bar{\phi}_1$,
\begin{eqnarray}
1&:&\gamma^o_{-1}=-\frac{\Lambda}{6}\\
r^{-1}&:&(\Psi_2)^o_1=0.
\end{eqnarray}

$DU=2\textrm{Re}(\bar{\tau}\omega-\gamma)$,
\begin{eqnarray}
r&:&U^o_{-2}=\frac{\Lambda}{6}.
\end{eqnarray}

Next, this Maxwell equation $(D'-2\gamma-\rho')\phi_0=(\delta-2\tau)\phi_1+\sigma\phi_2$ gives,
\begin{eqnarray}
r^{-1}:\frac{\Lambda}{6}(\phi_0)^o_2=0.
\end{eqnarray}
For a non-zero cosmological constant, we have $(\phi_0)^o_2=0$.

Hence, $\phi_0=O(r^{-3})$. With this fall-off for $\phi_0$, we have the following peeling property for the Maxwell spinor \cite{Vee2017}, i.e. $\phi_n=O(r^{n-3})$, for $n=0,1,2$.

Finally, this Bianchi identity $(D'-4\gamma-\rho')\Psi_0-(\delta-4\tau+2\alpha')\Psi_1-3\sigma\Psi_2=k\bar{\phi}_1(\delta+2\alpha')\phi_0-k\bar{\phi}_2D\phi_0+2k\sigma\phi_1\bar{\phi}_1$ gives,
\begin{eqnarray}\label{Psi4}
r^{-3}:\frac{\Lambda}{6}(\Psi_0)^o_4=0.
\end{eqnarray}
For a non-zero cosmological constant, we have $(\Psi_0)^o_4=0$ \footnote{In the derivation of the asymptotic solutions with $\Lambda$ as detailed in the appendices of Refs. \cite{Vee2016,Vee2017}, this Bianchi identity is the fourth last equation to be solved, i.e. the 35th out of the 38 Newman-Penrose equations (or the 39th out of the 42 Newman-Penrose-Maxwell equations). As the focus there was to produce the mass-loss formula instead of analysing the peeling property, it was not at all obvious that Eq. (\ref{Psi4}) would give $(\Psi_0)^o_4=0$ [there, the trivial equation ``$0=0$'' was obtained, since $\Psi_0=O(r^{-5})$ was assumed].}.

Thus, $\Psi_0=O(r^{-5})$. With this fall-off for $\Psi_0$, we have the peeling property for the Weyl spinor \cite{Vee2016,Vee2017}, i.e. $\Psi_n=O(r^{n-5})$, for $n=0,1,2,3,4$.

\emph{Remark}: If the Maxwell spinor is zero, then the assumption $\rho^o_1=-1$ is not required, since $D\rho=\rho^2+\sigma\bar{\sigma}$ gives at order $r^{-2}$ [Eq.(\ref{Drho})] $:\rho^o_1(\rho^o_1+1)=0$, i.e. $\rho^o_1=-1$ because $\rho^o_1\neq0$ [otherwise this would not even reduce to vacuum (anti-)de Sitter spacetime which does have $\rho=-r^{-1}$]. With the Maxwell fields involved, then Eq. (\ref{Drho}) would become: $\rho^o_1(\rho^o_1+1)+k|(\phi_0)^o_1|^2=0$, which does not guarantee that the congruence of null geodesics near $\mathcal{I}$ behaving like the purely vacuum case, i.e. with uniform divergence.

Essentially, a non-zero $\Lambda$ leads to new leading order terms in $U=\Lambda r^2/6+O(r)$, $\rho'=-\Lambda r/6+O(1)$ and $\gamma=-\Lambda r/6+O(1)$, such that the $D'$ derivative, $\rho'$ and $\gamma$ (or the \th$'$ derivative \cite{Pen87}) in the Maxwell equation/Bianchi identity would effectively shift the term involving $(\phi_0)^o_2$ or $(\Psi_0)^o_4$ by an order of $r$. Since this is the only term at that order, then $(\phi_0)^o_2$ or $(\Psi_0)^o_4$ has to be zero.

\section{Asymptotically de Sitter spacetimes with axisymmetry}

We work out the conditions on the coordinate transformations which preserve the asymptotic form of the metric. Following the steps of Bondi et al. who first worked this out with zero cosmological constant \cite{Bondi62}, let the old coordinates be related to the new coordinates by:
\begin{eqnarray}
u&=&a^0(\bar{u},\bar{\theta})+a^1(\bar{u},\bar{\theta})\bar{r}^{-1}+O(\bar{r}^{-2})\\
r&=&K(\bar{u},\bar{\theta})\bar{r}+\rho^0(\bar{u},\bar{\theta})+\rho^1(\bar{u},\bar{\theta})\bar{r}^{-1}+O(\bar{r}^{-2})\\
\theta&=&g^0(\bar{u},\bar{\theta})+g^1(\bar{u},\bar{\theta})\bar{r}^{-1}+O(\bar{r}^{-2})\\
\phi&=&\bar{\phi}.
\end{eqnarray}
The transformation matrix has components $\displaystyle\Lambda^a_{\ \bar{a}}=\frac{\partial x^a}{\partial x^{\bar{a}}}$:
\begin{eqnarray}
\Lambda^a_{\ \bar{a}}&=&
\begin{pmatrix}
  \Lambda^u_{\ \bar{u}} & \Lambda^u_{\ \bar{r}} & \Lambda^u_{\ \bar{\theta}} & \Lambda^u_{\ \bar{\phi}} \\
  \Lambda^r_{\ \bar{u}} & \Lambda^r_{\ \bar{r}} & \Lambda^r_{\ \bar{\theta}} & \Lambda^r_{\ \bar{\phi}}\\
  \Lambda^\theta_{\ \bar{u}} & \Lambda^\theta_{\ \bar{r}} & \Lambda^\theta_{\ \bar{\theta}} & \Lambda^\theta_{\ \bar{\phi}}\\
	\Lambda^\phi_{\ \bar{u}} & \Lambda^\phi_{\ \bar{r}} & \Lambda^\phi_{\ \bar{\theta}} & \Lambda^\phi_{\ \bar{\phi}}
\end{pmatrix}\\
&=&
\begin{pmatrix}
  a^0_{\bar{u}}+O(\bar{r}^{-1}) & -a^1\bar{r}^{-2}+O(\bar{r}^{-3}) & a^0_{\bar{\theta}}+O(\bar{r}^{-1}) & 0 \\
  K_{\bar{u}}\bar{r}+O(1) & K+O(\bar{r}^{-2}) & K_{\bar{\theta}}\bar{r}+O(1) & 0 \\
  g^0_{\bar{u}}+O(\bar{r}^{-1}) & -g^1\bar{r}^{-2}+O(\bar{r}^{-3}) & g^0_{\bar{\theta}}+O(\bar{r}^{-1}) & 0 \\
	0 & 0 & 0 & 1
\end{pmatrix}.
\end{eqnarray}
The metric has components (see Ref. \cite{Vee2016}):
\begin{eqnarray}
g_{ab}&=&
\begin{pmatrix}
  -\frac{\Lambda}{3}r^2+O(1) & 1 & O(1) & 0 \\
  1 & 0 & 0 & 0\\
  O(1) & 0 & -e^{2\Lambda f(u,\theta)}r^2+O(r) & 0\\
	0 & 0 & 0 & -e^{-2\Lambda f(u,\theta)}r^2\sin^2{\theta}+O(r)
\end{pmatrix}.
\end{eqnarray}
In the new set of coordinates $\bar{u},\bar{r},\bar{\theta},\bar{\phi}$, the metric has components $\bar{g}_{\bar{a}\bar{b}}=\Lambda^{a}_{\ \bar{a}}\Lambda^b_{\ \bar{b}}g_{ab}$, and must have the form:
\begin{eqnarray}
\bar{g}_{\bar{a}\bar{b}}&=&
\begin{pmatrix}
  -\frac{\Lambda}{3}\bar{r}^2+O(1) & 1 & O(1) & 0 \\
  1 & 0 & 0 & 0\\
  O(1) & 0 & -e^{2\Lambda F(\bar{u},\bar{\theta})}\bar{r}^2+O(\bar{r}) & 0\\
	0 & 0 & 0 & -e^{-2\Lambda F(\bar{u},\bar{\theta})}\bar{r}^2\sin^2{\bar{\theta}}+O(\bar{r})
\end{pmatrix},
\end{eqnarray}
where $F(\bar{u},\bar{\theta})$ is $f(u,\theta)$ expressed in the new coordinates. This function $3f(u,\theta):=\int{\sigma^o(u,\theta)du}$ causes $\mathcal{I}$ to be non-conformally flat whenever $\sigma^o$ (the leading order term of $\sigma$) is non-zero, i.e. when an isolated gravitating system radiates gravitational waves \cite{Vee2016}. Only in the absence of outgoing gravitational waves would $\mathcal{I}$ be conformally flat.

To leading order, these components are:
\begin{eqnarray}
\bar{g}_{\bar{u}\bar{u}}&=&\Lambda^a_{\ \bar{u}}\Lambda^b_{\ \bar{u}}g_{ab}\\
&=&\Lambda^u_{\ \bar{u}}\Lambda^u_{\ \bar{u}}g_{uu}+2\Lambda^u_{\ \bar{u}}\Lambda^r_{\ \bar{u}}g_{ur}+2\Lambda^u_{\ \bar{u}}\Lambda^\theta_{\ \bar{u}}g_{u\theta}+\Lambda^\theta_{\ \bar{u}}\Lambda^\theta_{\ \bar{u}}g_{\theta\theta}\\
&=&-\left(\frac{\Lambda}{3}K^2(a^0_{\bar{u}})^2+K^2(g^0_{\bar{u}})^2e^{2\Lambda f}\right)\bar{r}^2+O(\bar{r})\\
\bar{g}_{\bar{r}\bar{r}}&=&\Lambda^a_{\ \bar{r}}\Lambda^b_{\ \bar{r}}g_{ab}\\
&=&\Lambda^u_{\ \bar{r}}\Lambda^u_{\ \bar{r}}g_{uu}+2\Lambda^u_{\ \bar{r}}\Lambda^r_{\ \bar{r}}g_{ur}+2\Lambda^u_{\ \bar{r}}\Lambda^\theta_{\ \bar{r}}g_{u\theta}+\Lambda^\theta_{\ \bar{r}}\Lambda^\theta_{\ \bar{r}}g_{\theta\theta}\\
&=&-\left(2Ka^1+\frac{\Lambda}{3}K^2(a^1)^2+K^2(g^1)^2e^{2\Lambda f}\right)\bar{r}^{-2}+O(\bar{r}^{-3})\\
\bar{g}_{\bar{\theta}\bar{\theta}}&=&\Lambda^a_{\ \bar{\theta}}\Lambda^b_{\ \bar{\theta}}g_{ab}\\
&=&\Lambda^u_{\ \bar{\theta}}\Lambda^u_{\ \bar{\theta}}g_{uu}+2\Lambda^u_{\ \bar{\theta}}\Lambda^r_{\ \bar{\theta}}g_{ur}+2\Lambda^u_{\ \bar{\theta}}\Lambda^\theta_{\ \bar{\theta}}g_{u\theta}+\Lambda^\theta_{\ \bar{\theta}}\Lambda^\theta_{\ \bar{\theta}}g_{\theta\theta}\\
&=&-\left(\frac{\Lambda}{3}K^2(a^0_{\bar{\theta}})^2+K^2(g^0_{\bar{\theta}})^2e^{2\Lambda f}\right)\bar{r}^2+O(\bar{r})\\
\bar{g}_{\bar{\phi}\bar{\phi}}&=&\Lambda^a_{\ \bar{\phi}}\Lambda^b_{\ \bar{\phi}}g_{ab}\\
&=&\Lambda^\phi_{\ \bar{\phi}}\Lambda^\phi_{\ \bar{\phi}}g_{\phi\phi}\\
&=&-K^2e^{-2\Lambda f}\sin^2{g^0}\ \bar{r}^2+O(\bar{r})\\
\bar{g}_{\bar{u}\bar{r}}&=&\Lambda^a_{\ \bar{u}}\Lambda^b_{\ \bar{r}}g_{ab}\\
&=&\Lambda^u_{\ \bar{u}}\Lambda^u_{\ \bar{r}}g_{uu}+\left(\Lambda^u_{\ \bar{u}}\Lambda^r_{\ \bar{r}}+\Lambda^r_{\ \bar{u}}\Lambda^u_{\ \bar{r}}\right)g_{ur}+\left(\Lambda^u_{\ \bar{u}}\Lambda^\theta_{\ \bar{r}}+\Lambda^\theta_{\ \bar{u}}\Lambda^u_{\ \bar{r}}\right)g_{u\theta}+\Lambda^\theta_{\ \bar{u}}\Lambda^\theta_{\ \bar{r}}g_{\theta\theta}\\
&=&Ka^0_{\bar{u}}+\frac{\Lambda}{3}K^2a^0_{\bar{u}}a^1+K^2g^0_{\bar{u}}g^1e^{2\Lambda f}+O(\bar{r}^{-1})\\
\bar{g}_{\bar{u}\bar{\theta}}&=&\Lambda^a_{\ \bar{u}}\Lambda^b_{\ \bar{\theta}}g_{ab}\\
&=&\Lambda^u_{\ \bar{u}}\Lambda^u_{\ \bar{\theta}}g_{uu}+\left(\Lambda^u_{\ \bar{u}}\Lambda^r_{\ \bar{\theta}}+\Lambda^r_{\ \bar{u}}\Lambda^u_{\ \bar{\theta}}\right)g_{ur}+\left(\Lambda^u_{\ \bar{u}}\Lambda^\theta_{\ \bar{\theta}}+\Lambda^\theta_{\ \bar{u}}\Lambda^u_{\ \bar{\theta}}\right)g_{u\theta}+\Lambda^\theta_{\ \bar{u}}\Lambda^\theta_{\ \bar{\theta}}g_{\theta\theta}\\
&=&-\left(\frac{\Lambda}{3}K^2a^0_{\bar{u}}a^0_{\bar{\theta}}+K^2g^0_{\bar{u}}g^0_{\bar{\theta}}e^{2\Lambda f}\right)\bar{r}^2+O(\bar{r})\\
\bar{g}_{\bar{r}\bar{\theta}}&=&\Lambda^a_{\ \bar{r}}\Lambda^b_{\ \bar{\theta}}g_{ab}\\
&=&\Lambda^u_{\ \bar{r}}\Lambda^u_{\ \bar{\theta}}g_{uu}+\left(\Lambda^u_{\ \bar{r}}\Lambda^r_{\ \bar{\theta}}+\Lambda^r_{\ \bar{r}}\Lambda^u_{\ \bar{\theta}}\right)g_{ur}+\left(\Lambda^u_{\ \bar{r}}\Lambda^\theta_{\ \bar{\theta}}+\Lambda^\theta_{\ \bar{r}}\Lambda^u_{\ \bar{\theta}}\right)g_{u\theta}+\Lambda^\theta_{\ \bar{r}}\Lambda^\theta_{\ \bar{\theta}}g_{\theta\theta}\\
&=&Ka^0_{\bar{\theta}}+\frac{\Lambda}{3}K^2a^0_{\bar{\theta}}a^1+K^2g^0_{\bar{\theta}}g^1e^{2\Lambda f}+O(\bar{r}^{-1})\\
\bar{g}_{\bar{u}\bar{\phi}}&=&\bar{g}_{\bar{r}\bar{\phi}}=\bar{g}_{\bar{\theta}\bar{\phi}}=0,
\end{eqnarray}

\newpage

giving the following conditions:
\begin{eqnarray}
\bar{g}_{\bar{u}\bar{u}}&:&\frac{\Lambda}{3}K^2(a^0_{\bar{u}})^2+K^2(g^0_{\bar{u}})^2e^{2\Lambda f}=\frac{\Lambda}{3}\label{con1}\\
\bar{g}_{\bar{\theta}\bar{\theta}}&:&\frac{\Lambda}{3}K^2(a^0_{\bar{\theta}})^2+K^2(g^0_{\bar{\theta}})^2e^{2\Lambda f}=e^{2\Lambda F}\\
\bar{g}_{\bar{\phi}\bar{\phi}}&:&K^2e^{-2\Lambda f}\sin^2{g^0}=e^{-2\Lambda F}\sin^2{\bar{\theta}}\\
\bar{g}_{\bar{u}\bar{\theta}}&:&e^{2\Lambda f}g^0_{\bar{u}}g^0_{\bar{\theta}}=-\frac{\Lambda}{3}a^0_{\bar{u}}a^0_{\bar{\theta}}\label{con2}\\
\bar{g}_{\bar{r}\bar{\theta}}&:&Ka^0_{\bar{\theta}}+\frac{\Lambda}{3}K^2a^0_{\bar{\theta}}a^1+K^2g^0_{\bar{\theta}}g^1e^{2\Lambda f}=0\\
\bar{g}_{\bar{u}\bar{r}}&:&Ka^0_{\bar{u}}+\frac{\Lambda}{3}K^2a^0_{\bar{u}}a^1+K^2g^0_{\bar{u}}g^1e^{2\Lambda f}=1\\
\bar{g}_{\bar{r}\bar{r}}&:&2Ka^1+\frac{\Lambda}{3}K^2(a^1)^2+K^2(g^1)^2e^{2\Lambda f}=0.
\end{eqnarray}
The last three equations determine $a^1$ and $g^1$, quantities of the next order. Only two of these three equations are independent \footnote{Well, the conditions $\bar{g}_{\bar{r}\bar{\theta}}$ and $\bar{g}_{\bar{u}\bar{r}}$ can be solved for $a^1$ and $g^1$. These may be plugged into $\bar{g}_{\bar{r}\bar{r}}$ to show that it is trivially satisfied.}.

\section{No gravitational radiation, \texorpdfstring{$f=0,F=0$}{f=0, F=0}}

\subsection{The equations to solve}

In the absence of gravitational radiation, $f=0, F=0$ give the following set of transformation equations:
\begin{align}
\frac{\Lambda}{3}K^2(a^0_{\bar{u}})^2+K^2(g^0_{\bar{u}})^2&=\frac{\Lambda}{3}\\
\frac{\Lambda}{3}K^2(a^0_{\bar{\theta}})^2+K^2(g^0_{\bar{\theta}})^2&=1\\
K^2\sin^2{g^0}&=\sin^2{\bar{\theta}}\\
g^0_{\bar{u}}g^0_{\bar{\theta}}&=-\frac{\Lambda}{3}a^0_{\bar{u}}a^0_{\bar{\theta}}
\end{align}
From the first two equations,
\begin{align}\label{1/K2}
\frac{1}{K^2}=(a^0_{\bar{u}})^2+\frac{3}{\Lambda}(g^0_{\bar{u}})^2=\frac{\Lambda}{3}(a^0_{\bar{\theta}})^2+(g^0_{\bar{\theta}})^2
\end{align}
As $g^0_{\theta}\neq0$ (from the existence of the identity transformation $\theta\rightarrow\theta$), we can write the fourth equation as $g^0_{\bar{u}}=-\Lambda a^0_{\bar{u}}a^0_{\bar{\theta}}/3g^0_{\bar{\theta}}$. Plugging this into Eq.\ (\ref{1/K2}) gives
\begin{align}
(a^0_{\bar{u}})^2\left(1+\frac{\Lambda(a^0_{\bar{\theta}})^2}{3(g^0_{\bar{\theta}})^2}\right)=(g^0_{\bar{\theta}})^2\left(1+\frac{\Lambda(a^0_{\bar{\theta}})^2}{3(g^0_{\bar{\theta}})^2}\right).
\end{align}
Since $1+\Lambda(a^0_{\bar{\theta}})^2/3(g^0_{\bar{\theta}})^2\neq0$, we have
\begin{align}\label{a1}
a^0_{\bar{u}}=\pm g^0_{\bar{\theta}}.
\end{align}
Plugging this into the fourth equation gives the other relationship
\begin{align}\label{a2}
a^0_{\bar{\theta}}=\mp\frac{3}{\Lambda}g^0_{\bar{u}}.
\end{align}
Thus, we find that $a^0_{\bar{u}}$ and $a^0_{\bar{\theta}}$ are expressible in terms of $g^0_{\bar{\theta}}$ and $g^0_{\bar{u}}$, so solving for $g^0$ would give the solution for $a^0$ as well. In fact, differentiating Eq.\ (\ref{a1}) with respect to $\bar{\theta}$ and Eq.\ (\ref{a2}) with respect to $\bar{u}$ yields
\begin{align}
a^0_{\bar{u}\bar{\theta}}=\pm g^0_{\bar{\theta}\bar{\theta}}=\mp\frac{3}{\Lambda}g^0_{\bar{u}\bar{u}},
\end{align}
i.e. $g^0$ satisfies the Laplace equation
\begin{align}
\frac{3}{\Lambda}g^0_{\bar{u}\bar{u}}+g^0_{\bar{\theta}\bar{\theta}}=0.
\end{align}

The other equation that $g^0$ has to satisfy is the third equation, which becomes
\begin{align}
\frac{1}{K^2}=\frac{\sin^2{g^0}}{\sin^2{\bar{\theta}}}=\frac{3}{\Lambda}(g^0_{\bar{u}})^2+(g^0_{\bar{\theta}})^2.
\end{align}

\subsection{Solving the equations}

Let us rename $g^0(\bar{u},\bar{\theta})$ as $G(\bar{u},\bar{\theta})$. The two equations to solve for $G(\bar{u},\bar{\theta})$ are
\begin{gather}
\frac{3}{\Lambda}G_{\bar{u}\bar{u}}+G_{\bar{\theta}\bar{\theta}}=0\label{firstonetosolve}\\
\frac{\sin^2{G}}{\sin^2{\bar{\theta}}}=\frac{3}{\Lambda}(G_{\bar{u}})^2+(G_{\bar{\theta}})^2.\label{otheronetosolve}
\end{gather}
Under the following change of variables $x=\bar{\theta},y=\bar{u}\sqrt{\Lambda/3},z=x+iy,\bar{z}=x-iy$, we can convert $G(\bar{u},\bar{\theta})$ into $G(z,\bar{z})$. Well,
\begin{align}
G_{\bar{u}}&=\sqrt{\frac{\Lambda}{3}}G_y\\
G_{\bar{\theta}}&=G_x\\
G_x&=G_z+G_{\bar{z}}\\
G_y&=i(G_z-G_{\bar{z}}),
\end{align}
so
\begin{align}
\frac{3}{\Lambda}G_{\bar{u}\bar{u}}&=G_{yy}=-G_{zz}-G_{\bar{z}\bar{z}}+2G_{z\bar{z}}\\
G_{\bar{\theta}\bar{\theta}}&=G_{xx}=G_{zz}+G_{\bar{z}\bar{z}}+2G_{z\bar{z}},\\
\frac{3}{\Lambda}(G_{\bar{u}})^2&=(G_y)^2=-(G_z)^2-(G_{\bar{z}})^2+2G_zG_{\bar{z}}\\
(G_{\bar{\theta}})^2&=(G_x)^2=(G_z)^2+(G_{\bar{z}})^2+2G_zG_{\bar{z}},
\end{align}
and the two differential equations become
\begin{align}
\frac{\partial^2G(z,\bar{z})}{\partial z\partial\bar{z}}&=0\\
\frac{\sin^2{G(z,\bar{z})}}{\sin^2{(\textrm{Re}(z))}}&=4\frac{\partial G(z,\bar{z})}{\partial z}\frac{\partial G(z,\bar{z})}{\partial\bar{z}}.
\end{align}
The former can be integrated twice to produce
\begin{align}
G(z,\bar{z})=\frac{1}{2}(\psi(z)+\chi(\bar{z})),
\end{align}
where $\psi(z)$ and $\chi(\bar{z})$ are arbitrary functions of $z$ and $\bar{z}$, respectively. Plugging this solution into the latter yields
\begin{align}
\frac{\sin^2{((\psi(z)+\chi(\bar{z}))/2)}}{\sin^2{((z+\bar{z})/2)}}=\frac{d\psi(z)}{dz}\frac{d\chi(\bar{z})}{d\bar{z}}.
\end{align}
Now, we only consider real solutions $G(z,\bar{z})=\overline{G(z,\bar{z})}$, so $\psi(z)+\chi(\bar{z})=\overline{\chi(\bar{z})}+\overline{\psi(z)}$. Hence $\chi(\bar{z})=\overline{\psi(z)}$, i.e. $G(z,\bar{z})=\textrm{Re}(\psi(z))$ and also $\frac{d\chi(\bar{z})}{d\bar{z}}=\frac{d\overline{\psi(z)}}{d\bar{z}}=\overline{\frac{d\psi(z)}{dz}}$.

We may express the differential equation that we need to solve as
\begin{align}
\sin{(G(z,\bar{z}))}=\pm A(z)\overline{A(z)}\sin x,
\end{align}
where $A(z)=\sqrt{d\psi(z)/dz}=\sqrt{\psi'(z)}$. Differentiating with respect to $x$ gives
\begin{align}
G_x\cos{G}=\pm(2\textrm{Re}(A\overline{A'})\sin{x}+|A|^2\cos{x}).
\end{align}
Differentiating with respect to $y$ gives
\begin{align}
G_y\cos{G}=\pm2\textrm{Im}(A\overline{A'})\sin{x}.
\end{align}
The sum of squares of these two equations is
\begin{align}
((G_x)^2+(G_y)^2)\cos^2{G}=4|AA'|^2\sin^2{x}+|A|^4\cos^2{x}+4|A|^4\textrm{Re}\left(\frac{A'}{A}\right)\sin{x}\cos{x}.
\end{align}
Note that $(G_x)^2+(G_y)^2=|\psi'|^2$, $\cos^2{G}=1-\sin^2{G}=1-|\psi'|^2\sin^2{x}$, $4|AA'|^2=|\psi''|^2$, $|A|^4=|\psi'|^2$, $4|A|^4\textrm{Re}(A'/A)=2|\psi'|^2\textrm{Re}(\psi''/\psi')$. Therefore,
\begin{align}
(|\psi''|^2+|\psi'|^4-|\psi'|^2)\sin{x}+2|\psi'|^2\textrm{Re}\left(\frac{\psi''}{\psi'}\right)\cos{x}=0.
\end{align}
Let $p=|\psi''|^2+|\psi'|^4-|\psi'|^2$ and $q=2|\psi'|^2\textrm{Re}(\psi''/\psi')$. We can express $p\sin{x}+q\cos{x}$ as $R\sin(x+\alpha)$ with $R=\sqrt{p^2+q^2}$ and $\alpha=\arctan{(q/p)}$, by letting $p=R\cos{\alpha}$ and $q=R\sin{\alpha}$. Thus,
\begin{align}
R\sin{(x+\alpha)}=0,
\end{align}
with two cases to solve, viz. $R=0$ or $\alpha=n\pi-x,n\in\Z$.

\subsubsection{\texorpdfstring{$R=0$}{R=0}}

Since $p$ and $q$ are real, then $p^2+q^2=R^2=0$ is a circle with radius $0$, i.e. $p=0$ and $q=0$. From the latter, we get $|\psi'|^2=0$ or $\textrm{Re}(\psi''/\psi')=0$. If $|\psi'|^2=0$, then $\psi'=0$ (which satisfies $p=0$) and $\sin{G}=\pm|\psi'|\sin{x}$ implies $G=k\pi, k\in\Z$. On the other hand, $|\psi'|\neq0$ and $\textrm{Re}(\psi''/\psi')=0$ imply that $\psi''/\psi'$ is a constant, because $\psi''(z)/\psi'(z)=-\overline{\psi''(z)/\psi'(z)}$ is such that LHS is a function of $z$ but RHS is a function of $\bar{z}$. Moreover, $\psi''/\psi'=ic$, with $c\in\R$ because $\textrm{Re}(\psi''/\psi')=0$. Integrating once gives $\psi'(z)=we^{icz}$, where $w\in\C$ is a constant. Plugging this into $p=0$, we get
\begin{align}
|\psi'|^2(c^2+|\psi'|^2-1)=0,
\end{align}
i.e. $|\psi'|^2=0$ or $|\psi'|^2=1-c^2$. For the former, we already know that this gives $G=k\pi,k\in\Z$. So from now on, we only consider $|\psi'|\neq0$. Then, $|\psi'|^2=1-c^2$ gives $|w|^2e^{-2cy}=1-c^2$. Since RHS is a constant, then LHS cannot depend on $y$, i.e. $c=0$, giving $|w|=1$. Thus, $\psi'(z)=e^{i\phi}$, $\phi\in\R$ is a constant, $-\pi<\phi\leq\pi$, so
\begin{align}
\psi(z)=ze^{i\phi}+z_0,
\end{align}
where $z_0\in\C$ is a constant. From $\sin{G}=\pm|\psi'|\sin{x}$ and $G(x,y)=\textrm{Re}(\psi(x+iy))$, we have
\begin{align}
\sin{(\textrm{Re}((x+iy)e^{i\phi}+z_0))}=\sin(\pm x).
\end{align}
Since RHS does not depend on $y$, then $e^{i\phi}$ must be real, i.e. $\phi=0$ or $\phi=\pi$, so that LHS also does not depend on $y$. This gives $\sin{(\pm x+\textrm{Re}(z_0))}=\sin(\pm x)$, i.e. $\textrm{Re}(z_0)=k\pi,k\in\Z$. Ergo,
\begin{align}
G(\bar{u},\bar{\theta})=\pm\bar{\theta}+k\pi,k\in\Z.
\end{align}

\subsubsection{\texorpdfstring{$\sin{(x+\alpha)}=0\textrm{, i.e. }\alpha=n\pi-x,n\in\Z$}{sin(x+a)=0}}

The equations to solve are
\begin{align}
p&=|\psi''|^2+|\psi'|^4-|\psi'|^2=R\cos{\alpha}=\pm R\cos{x}\label{plah}\\
q&=2|\psi'|^2\textrm{Re}(\psi''/\psi')=R\sin{\alpha}=\mp R\sin{x}.\label{qlah}
\end{align}
We first consider the latter:
\begin{align}
\frac{\psi(z)''}{\psi'(z)}+\overline{\left(\frac{\psi(z)''}{\psi'(z)}\right)}=\frac{1}{2i}S(z,\bar{z})(e^{i(z+\bar{z})/2}-e^{-i(z+\bar{z})/2}),\label{conpareconstants}
\end{align}
where $S(z,\bar{z})=\mp R(z,\bar{z})/|\psi'(z)|^2$, and $|\psi'|\neq0$. Note that $\overline{\psi''(z)/\psi'(z)}$ is a function of $\bar{z}$ only, so differentiating with respect to $z$ gives
\begin{align}\label{tobesplitup}
2i\frac{d}{dz}\left(\frac{\psi''(z)}{\psi'(z)}\right)=\left(S_z+\frac{i}{2}S\right)e^{i(z+\bar{z})/2}-\left(S_z-\frac{i}{2}S\right)e^{-i(z+\bar{z})/2}.
\end{align}
Let
\begin{align}
\left(S_z+\frac{i}{2}S\right)e^{i(z+\bar{z})/2}&=U(z)+\eta(z,\bar{z})\label{someone}\\
\left(S_z-\frac{i}{2}S\right)e^{-i(z+\bar{z})/2}&=V(z)+\zeta(z,\bar{z}).\label{sometwo}
\end{align}
This decomposition isolates the parts $U(z)$ and $V(z)$ that are purely $z$-dependent, with the remainders $\eta(z,\bar{z})$ and $\zeta(z,\bar{z})$ dependent on both $z$ and $\bar{z}$. Note that LHS of Eq.\ (\ref{tobesplitup}) is purely a function of $z$, which demands the same for RHS. Hence, $\eta(z,\bar{z})$ would cancel out $\zeta(z,\bar{z})$, i.e. $\eta(z,\bar{z})=\zeta(z,\bar{z})$. If we move the exponential factors in Eqs.\ (\ref{someone})-(\ref{sometwo}) to RHS:
\begin{align}
S_z+\frac{i}{2}S&=(U(z)+\eta(z,\bar{z}))e^{-i(z+\bar{z})/2}\label{ODE1}\\
S_z-\frac{i}{2}S&=(V(z)+\eta(z,\bar{z}))e^{i(z+\bar{z})/2},\label{ODE2}
\end{align}
then Eqs.\ (\ref{ODE1})-(\ref{ODE2}) form a system of two linear inhomogeneous ordinary differential equations (ODE) for $S(z,\bar{z})$ with constant coefficients. Typically if there is one linear ODE with inhomogeneous terms, the latter is specified and we then solve this ODE. Here, however, the inhomogeneous terms are unknown. Nevertheless, since $S(z,\bar{z})$ has to simultaneously satisfy \emph{two} such ODE, it turns out that we are able to systematically determine not only $S(z,\bar{z})$, but also the inhomogeneous unknown functions $U(z),V(z),\eta(z,\bar{z})$.

In solving a linear inhomogeneous ODE with constant coefficients, the first thing to do is to get the complementary function for the corresponding homogeneous ODE. These are
\begin{align}
S(z,\bar{z})&=S_1(\bar{z})e^{-iz/2}\\
S(z,\bar{z})&=S_2(\bar{z})e^{iz/2},
\end{align}
for Eqs.\ (\ref{ODE1})-(\ref{ODE2}), respectively, where $S_1(\bar{z})$ and $S_2(\bar{z})$ are arbitrary functions of $\bar{z}$. The common solution for $S(z,\bar{z})$ that satisfies both homogeneous ODE is if $S_1(\bar{z})=S_2(\bar{z})=0$, i.e. $S(z,\bar{z})=0$ is the complementary function for this pair of ODE. Let us deal with the inhomogeneous term $U(z)e^{-i(z+\bar{z})/2}$ in Eq.\ (\ref{ODE1}), i.e.
\begin{align}
S_z+\frac{i}{2}S=Ue^{-i(z+\bar{z})/2},
\end{align}
with $\eta=0$. The other ODE, when $\eta=0$, is
\begin{align}
S_z-\frac{i}{2}S=Ve^{i(z+\bar{z})/2}.
\end{align}
Adding them gives
\begin{align}
2S_z=Ue^{-i(z+\bar{z})/2}+Ve^{i(z+\bar{z})/2}.\label{asd}
\end{align}
Subtracting one from the other gives
\begin{align}
iS&=Ue^{-i(z+\bar{z})/2}-Ve^{i(z+\bar{z})/2}.
\end{align}
Differentiating with respect to $z$,
\begin{align}
2S_z&=-(2iU_z+U)e^{-i(z+\bar{z})/2}+(2iV_z-V)e^{i(z+\bar{z})/2},
\end{align}
and so with Eq.\ (\ref{asd}), we have
\begin{align}
iU_z+U=(iV_z-V)e^{i(z+\bar{z})}.
\end{align}
Since $U(z)$ and $V(z)$ are functions of $z$, then $iV_z=V$ and $iU_z=-U$, i.e. $U(z)=is_1e^{iz}$ and $V(z)=-is_2e^{-iz}$, where $s_1,s_2\in\C$ are constants. Therefore, $S(z,\bar{z})=s_1e^{i(z-\bar{z})/2}+s_2e^{-i(z-\bar{z})/2}$ is the general solution to the pair of ODE:
\begin{align}
S_z+\frac{i}{2}S&=is_1e^{i(z-\bar{z})/2}\\
S_z-\frac{i}{2}S&=-is_2e^{-i(z-\bar{z})/2}.
\end{align}
Note that $S(z,\bar{z})=s_1e^{-y}+s_2e^{y}$ is real. Since $e^y$ and $e^{-y}$ are linearly independent, then $s_1$ and $s_2$ are also real. Here, we solved for $U$ and $V$ simultaneously. The same result would be reached, if they are solved individually.

This leaves us with the inhomogeneous term $\eta(z,\bar{z})$. Consider the pair of ODE:
\begin{align}
S_z+\frac{i}{2}S&=\eta e^{-i(z+\bar{z})/2}\\
S_z-\frac{i}{2}S&=\eta e^{i(z+\bar{z})/2}.
\end{align}
Dividing to eliminate $\eta$ and rearranging, we have
\begin{align}
\frac{1}{S}\frac{dS}{dz}&=-\frac{\cos{((z+\bar{z})/2)}}{2\sin{((z+\bar{z})/2)}}\\
\int{\frac{1}{S}dS}&=-\int{\frac{\cos{((z+\bar{z})/2)}}{2\sin{((z+\bar{z})/2)}}dz}\\
S(z,\bar{z})&=S_3(\bar{z})\csc{\left(\frac{z+\bar{z}}{2}\right)},
\end{align}
where $S_3(\bar{z})$ is an arbitrary function of $\bar{z}$. Note that since $S$ is real, then $S_3(\bar{z})$ must be real, i.e. $S_3(\bar{z})=s_3$ is a real constant. So, $S(z,\bar{z})=s_3\csc{((z+\bar{z})/2)}$. This gives $\eta(z,\bar{z})=-(s_3/2)\csc^2{((z+\bar{z})/2)}$. Ergo,
\begin{align}
S(z,\bar{z})=s_1e^{i(z-\bar{z})/2}+s_2e^{-i(z-\bar{z})/2}+s_3\csc{\left(\frac{z+\bar{z}}{2}\right)},
\end{align}
where $s_1,s_2,s_3\in\R$ are constants, is the general solution to the pair of ODE
\begin{align}
S_z+\frac{i}{2}S&=is_1e^{i(z-\bar{z})/2}-\frac{1}{2}s_3\csc^2{\left(\frac{z+\bar{z}}{2}\right)}e^{-i(z+\bar{z})/2}\\
S_z-\frac{i}{2}S&=-is_2e^{-i(z-\bar{z})/2}-\frac{1}{2}s_3\csc^2{\left(\frac{z+\bar{z}}{2}\right)}e^{i(z+\bar{z})/2}.
\end{align}

Now, from Eq.\ (\ref{tobesplitup}), we have
\begin{align}
2i\frac{d}{dz}\left(\frac{\psi''(z)}{\psi'(z)}\right)&=U(z)-V(z)\\
&=is_1e^{iz}+is_2e^{-iz},
\end{align}
so \emph{indeed LHS and RHS are purely functions of $z$}. Integrating once gives
\begin{align}\label{someLHSlah}
\frac{1}{\psi'}\frac{d\psi'}{dz}=\frac{1}{2i}(s_1e^{iz}-s_2e^{-iz})+\varepsilon,
\end{align}
where $\varepsilon\in\C$ is a constant. Integrating again gives
\begin{align}
\psi'(z)=\mu\exp{\left[\varepsilon z-\frac{1}{2}\left(s_1e^{iz}+s_2e^{-iz}\right)\right]},
\end{align}
where $\mu\in\C$ is a constant. Note that from Eq.\ (\ref{conpareconstants}),
\begin{gather}
2\textrm{Re}\left(\frac{\psi''(z)}{\psi'(z)}\right)=\frac{1}{2i}\left(s_1e^{i(z-\bar{z})/2}+s_2e^{-i(z-\bar{z})/2}+s_3\csc{\left(\frac{z+\bar{z}}{2}\right)}\right)(e^{i(z+\bar{z})/2}-e^{-i(z+\bar{z})/2}).
\end{gather}
By expanding and grouping up terms, RHS can be expressed as
\begin{align}
\textrm{Re}\left(\frac{1}{i}(s_1e^{iz}-s_2e^{-iz})\right)+s_3.
\end{align}
With this and LHS directly obtainable from Eq.\ (\ref{someLHSlah}), we get
\begin{gather}
\textrm{Re}\left(\frac{1}{i}(s_1e^{iz}-s_2e^{-iz})+2\varepsilon\right)=\textrm{Re}\left(\frac{1}{i}(s_1e^{iz}-s_2e^{-iz})\right)+s_3\\
2\textrm{Re}\left(\varepsilon\right)=s_3.
\end{gather}
This concludes Eq.\ (\ref{qlah}).

The other equation to solve is Eq.\ (\ref{plah}),
\begin{align}
|\psi'|^2\left(\left|\frac{\psi''}{\psi'}\right|^2+|\psi'|^2-1\right)&=R\cos{(n\pi-x)},n\in\Z\\
\left|\frac{\psi''}{\psi'}\right|^2+|\psi'|^2-1&=-S\cos{x},
\end{align}
where $S(z,\bar{z})=\mp R(z,\bar{z})/|\psi'|^2$ and $|\psi'|\neq0$. Well,
\begin{align}
\left|\frac{\psi''}{\psi'}\right|^2&=\left[\frac{1}{2i}(s_1e^{iz}-s_2e^{-iz})+\varepsilon\right]\left[-\frac{1}{2i}(s_1e^{-i\bar{z}}-s_2e^{i\bar{z}})+\bar{\varepsilon}\right]\\
|\psi'|^2&=|\mu|^2\exp{\left[2\textrm{Re}(\varepsilon z)-(s_1e^{-y}+s_2e^y)\cos{x}\right]}\\
S&=s_1e^{-y}+s_2e^y+2\textrm{Re}(\varepsilon)\csc{x}.
\end{align}
When $|\psi''/\psi'|^2+|\psi'|^2-1=-S\cos{x}$ is evaluated, we find that there is only one appearance of the double exponent $\exp[-(s_1e^{-y}+s_2e^y)\cos{x}]$, which depends on $x$ and $y$. Eliminating the $x$-dependence requires $s_1e^{-y}+s_2e^y$ to be zero, i.e. $s_1=-s_2e^{2y}$. Since $s_1$ and $s_2$ are constants, then $s_2=0$ and $s_1=0$. Consequently, $|\psi''/\psi'|^2=|\varepsilon|^2$, $|\psi'|^2=|\mu|^2e^{2\textrm{Re}(\varepsilon z)}$, $S=2\textrm{Re}(\varepsilon)\csc{x}$. So,
\begin{align}
|\varepsilon|^2+|\mu|^2e^{2\textrm{Re}(\varepsilon z)}-1=-2\textrm{Re}(\varepsilon)\cot{x}.
\end{align}
As there is no $y$-dependence on RHS, then the same must be true for LHS, i.e. $\text{Im}(\varepsilon)=0$, giving
\begin{align}
\textrm{Re}(\varepsilon)^2+|\mu|^2e^{2x\textrm{Re}(\varepsilon)}-1=-2\textrm{Re}(\varepsilon)\cot{x}.
\end{align}
Since $e^{2x\textrm{Re}(\varepsilon)}$ and $\cot{x}$ are linearly independent, they must both vanish. The vanishing of the latter demands $\textrm{Re}(\varepsilon)=0$, which automatically eliminates the former's $x$-dependence. Thus, $|\mu|^2=1$, i.e. $|\psi'|=1$. Earlier, we knew that this implies $G(\bar{u},\bar{\theta})=\pm\bar{\theta}+k\pi,k\in\Z$. Ergo, the complete solution to
\begin{gather}
\frac{3}{\Lambda}G_{\bar{u}\bar{u}}+G_{\bar{\theta}\bar{\theta}}=0\\
\frac{\sin^2{G}}{\sin^2{\bar{\theta}}}=\frac{3}{\Lambda}(G_{\bar{u}})^2+(G_{\bar{\theta}})^2,
\end{gather}
are
\begin{align}
G(\bar{u},\bar{\theta})=m\bar{\theta}+k\pi,m\in\{-1,0,-1\},k\in\Z.
\end{align}

The only valid physical solution is $G(\bar{\theta})=\bar{\theta}$, which is the identity transformation. The solution where $m=0$ is nonsense, because it makes the $\bar{\theta}$ coordinate vanish. The other solution $m=-1$ is a discrete swap from $\bar{\theta}$ to $-\bar{\theta}$. The remaining freedom $k\neq0$ merely shifts $\bar{\theta}$ by multiples of $\pi$. After all these calculations, the implication is there is \emph{no room for supertranslations}.

\section{Discussion}

Although it is well-known that the peeling property holds for asymptotically simple spacetimes by studying its conformal structure \cite{Pen65,Pen88}, the original physical spacetime approach by Newman and Penrose requires $\Psi_0=O(r^{-5})$ to be stated as an assumption \cite{newpen62}. Indeed, this leads to the deduction of the fall-offs of all other quantities (for the asymptotically flat case \cite{newpen62}) and the peeling property of the Weyl spinor (for the asymptotically flat case \cite{newpen62} as well as with a cosmological constant \cite{Vee2016}).

The point in this paper, based exclusively on the physical spacetime is: \emph{Even though the fall-offs of the spin coefficients [Eqs. (\ref{spincoe1})-(\ref{spincoe2})], the unknown functions in the null tetrad [Eqs. (\ref{U})-(\ref{xi})], and the vanishing of the Weyl spinor on $\mathcal{I}$ [Eqs. (\ref{Weyl1})-(\ref{Weyl2})] are enunciated as an ansatz, it appears that one can deduce that $\Psi_0=O(r^{-5})$ only when there is a non-zero cosmological constant.} Hence ironically, whilst these fall-offs together with the vanishing of the Weyl spinor on $\mathcal{I}$ are consistent with admitting a smooth conformal compactifiability \cite{Szabados}, it seems that these are insufficient to guarantee the peeling property if the cosmological constant is zero. This is in contrast to the fully conformal approach with explicit use of spinors that ensures the peeling property for asymptotically simple spacetimes \cite{Pen65,Pen88}.

A result that is strikingly comparable (but not equivalent) to what we have shown here was recently arrived at by Ortaggio and Pravdov\'a \cite{Ortaggio1}, in an exclusive study of the peeling behaviour of the Weyl tensor in an $n$-dimensional spacetime (where $n=4$ is a special case). They found that for our four-dimensional spacetime, the presence of a \emph{non-zero cosmological constant} would guarantee the usual peeling property [see Eq. (1) in Ref. \cite{Ortaggio1}] based on the assumptions that $\Psi_0$ falls off faster than $1/r^2$ as well as the optical matrix of the outgoing null geodesics being asymptotically non-singular and expanding. Unlike here, they did not make any further stipulation on the fall-offs of the Ricci rotation coefficients (i.e. the spin coefficients) nor the fall-offs of the unknown functions in the null tetrad (although for some cases, there is a need to additionally say something about the fall-offs of $\Psi_1$ and $\Psi_2$ --- see the last paragraph of Section II A in Ref. \cite{Ortaggio1}). This is in a similar spirit to the original Newton-Penrose approach of making a single assumption of $\Psi_0=O(r^{-5})$ to deduce the fall-offs of everything else \cite{newpen62}, but represents an improvement since a weaker assumption is made and it also includes the case with $\Lambda\neq0$. The case for $\Lambda=0$ however, allows for more possibilities apart from the usual peeling [see Eq. (2) in Ref. \cite{Ortaggio1}]. A treatment for the Maxwell case is given in Ref. \cite{Ortaggio2}.

Incidentally, the peeling property of the Weyl spinor in a universe with a cosmological constant has been shown explicitly within a linearised setup by Bishop, who worked with the Bondi-Sachs framework \cite{gracos1}. More recently, an intriguing new boundary condition was proposed by Xie and Zhang when there is a cosmological constant, which also leads to the usual peeling property \cite{Zhang}. In particular, they showed that $\Psi_4$ is of order $O(r^{-1})$ in spite of the fact that there is no Bondi news in their metric. Their new boundary condition corresponds to our $X^\mu$ in Eq. (\ref{X}) having a fall-off of order $O(1)$ instead of $O(r^{-1})$ \footnote{I thank Xiao Zhang for explicitly pointing this out to me, in terms of the Newman-Unti null tetrad that I have employed.}. This however, would lead to an asymptotic form of the metric that cannot be derived from the conformal approach. More specifically, their metric, when compared to Eq. (4.18) in the work of Szabados and Tod (who produced the asymptotic fall-offs for the physical metric based on the conformal approach \cite{Szabados}), differs in the $g_{uu}$ as well as the $g_{u\theta}$ components --- and \emph{cannot be obtained from Eq. (4.18) of Ref. \cite{Szabados}}. Nevertheless, we live in physical space and the property that there is a smooth conformal extension at infinity (which is the basis for such a conformal compactification by Penrose \cite{Pen65}) is an assumption --- albeit a highly useful one and is the mainstream compactification method in general relativity. In fact, there are other ways to compactify spacetime, for instance one which preserves geodesics, viz. projective compactification \cite{Rod} \footnote{Private communication with J{\"o}rg Frauendiener.}.

As this study originated from the goal of extending the Bondi mass and the mass loss of an isolated gravitating system due to energy carried away by gravitational waves to include a cosmological constant \cite{Vee2016}, we wish to emphasise here that the explicit manifestation of the peeling property depends on the choices of: 1) coordinates, and 2) null tetrad. It has been shown and explained in Ref. \cite{Vee2017} that a different null tetrad (using the same Bondi-Sachs coordinates) would not exhibit the peeling property. There, the Weyl and Maxwell spinors were given in the usual Newman-Unti null tetrad as well as the Szabados-Tod null tetrad, which are related by a boost transformation. The peeling property is displayed in the former, but not in the latter. In general (see Appendix A in Ref. \cite{Vee2017}), a boost transformation would affect the fall-offs of various quantities. Also, a proof of the peeling property given in Ref. \cite{Pen88} requires that $r$ is an affine parameter of the outgoing null generators tangent to the null tetrad vector $\vec{l}$, so if $r$ and $\vec{l}$ are not compatible (as in the case for the Szabados-Tod null tetrad), then the peeling property is not depicted.

Alternatively, a different choice of coordinates (one which is not of the Bondi-Sachs type) may not portray the peeling property. A work by Ashtekar et al. \cite{ash3}, who explored the linearised theory with a cosmological constant, adopted coordinates describing the future Poincar\'e patch. They found that with such a coordinate system, the usual expansion in powers of $1/$``$r$'' does not quite work, and that ``$r$''$\rightarrow\infty$ gives the past cosmological horizon instead of future null infinity $\mathcal{I}$. Note that the ``$r$'' in Ref. \cite{ash3} is not the same as our $r$ here and in Refs. \cite{Vee2016,Vee2017}, which \emph{does} give $\mathcal{I}$ in the limit where $r\rightarrow\infty$ \footnote{Recall that our $r$ is an affine parameter of the null generators tangent to the null tetrad vector $\vec{l}$, which is an outgoing null vector. Then, $r\rightarrow\infty$ goes to infinity along a null direction, i.e. approaches future null infinity $\mathcal{I}$.}.

These different choices of the coordinate systems would turn out to lead to purportedly different results, in terms of the energy carried away by gravitational waves in a universe with $\Lambda$. According to the exact result based on the full Einstein theory, Ref. \cite{Vee2016} showed that in the approximation where $\Lambda$ is tiny, then the correction to the mass-loss formula due to energy carried away by gravitational waves is in \emph{positive integer powers of $\Lambda$}. This is in agreement with the linearised treatment carried out by Bishop \cite{gracos1}. The result obtained in Ref. \cite{ash3} however, is a correction in \emph{powers of $\sqrt{\Lambda}$}. Interestingly, work done by Date and Hoque within the linearised theory considered two different coordinate systems \cite{gracos2}. With the ``Fermi normal coordinates'', the correction is in powers of $\Lambda$ --- in accordance to the results of Refs. \cite{Vee2016,Vee2017,gracos1}. On the other hand, they found that the use of a ``conformal chart'' gives a correction in powers of $\sqrt{\Lambda}$.

\section*{Acknowledgements}
V.-L. Saw was supported by the University of Otago Doctoral Scholarship, when the work on the peeling property was carried out in University of Otago, New Zealand.

\bibliographystyle{spphys}
\bibliography{Citation}

\begin{thebibliography}{10}
\providecommand{\url}[1]{{#1}}
\providecommand{\urlprefix}{URL }
\expandafter\ifx\csname urlstyle\endcsname\relax
  \providecommand{\doi}[1]{DOI \discretionary{}{}{}#1}\else
  \providecommand{\doi}{DOI \discretionary{}{}{}\begingroup
  \urlstyle{rm}\Url}\fi

\bibitem{Sachs61}
R.~Sachs, Proceedings of the Royal Society of London A: Mathematical, Physical
  and Engineering Sciences \textbf{264}(1318), 309 (1961).
\newblock \doi{10.1098/rspa.1961.0202}.
\newblock
  \urlprefix\url{http://rspa.royalsocietypublishing.org/content/264/1318/309}

\bibitem{Bondi62}
H.~Bondi, M.G.J. van~der Burg, A.W.K. Metzner, Proceedings of the Royal Society
  of London A: Mathematical, Physical and Engineering Sciences
  \textbf{269}(1336), 21 (1962).
\newblock \doi{10.1098/rspa.1962.0161}.
\newblock
  \urlprefix\url{http://rspa.royalsocietypublishing.org/content/269/1336/21}

\bibitem{Sachs62}
R.K. Sachs, Proceedings of the Royal Society of London A: Mathematical,
  Physical and Engineering Sciences \textbf{270}(1340), 103 (1962).
\newblock \doi{10.1098/rspa.1962.0206}.
\newblock
  \urlprefix\url{http://rspa.royalsocietypublishing.org/content/270/1340/103}

\bibitem{newpen62}
E.T. Newman, R.~Penrose, Journal of Mathematical Physics \textbf{3}(3), 566
  (1962).
\newblock \doi{http://dx.doi.org/10.1063/1.1724257}.
\newblock
  \urlprefix\url{http://scitation.aip.org/content/aip/journal/jmp/3/3/10.1063/1.1724257}

\bibitem{newunti62}
E.T. Newman, T.W.J. Unti, Journal of Mathematical Physics \textbf{3}(5), 891
  (1962).
\newblock \doi{http://dx.doi.org/10.1063/1.1724303}.
\newblock
  \urlprefix\url{http://scitation.aip.org/content/aip/journal/jmp/3/5/10.1063/1.1724303}

\bibitem{Vee2016}
V.-L. Saw, Phys. Rev. D \textbf{94}, 104004 (2016).
\newblock \doi{10.1103/PhysRevD.94.104004}.
\newblock \urlprefix\url{https://link.aps.org/doi/10.1103/PhysRevD.94.104004}

\bibitem{Vee2017}
V.-L. Saw, Phys. Rev. D \textbf{95}, 084038 (2017).
\newblock \doi{10.1103/PhysRevD.95.084038}.
\newblock \urlprefix\url{https://link.aps.org/doi/10.1103/PhysRevD.95.084038}

\bibitem{Vee2017b}
V.-L. Saw, Modern Physics Letters A \textbf{32}, 1730020 (2017).
\newblock \doi{10.1142/S0217732317300208}.
\newblock
  \urlprefix\url{http://www.worldscientific.com/doi/abs/10.1142/S0217732317300208}

\bibitem{Vee2017d}
V.-L. Saw, International Journal of Modern Physics D \textbf{27}(01), 1730027
  (2018).
\newblock \doi{10.1142/S0218271817300270}.
\newblock
  \urlprefix\url{http://www.worldscientific.com/doi/abs/10.1142/S0218271817300270}

\bibitem{Vee2018}
V.-L. Saw, Phys. Rev. D \textbf{97}, 084017 (2018).
\newblock \doi{10.1103/PhysRevD.97.084017}.
\newblock \urlprefix\url{https://link.aps.org/doi/10.1103/PhysRevD.97.084017}

\bibitem{Note1}
In the latest development, we found that gravitational waves radiated by a
  \protect \emph {quadrupole source} must carry a \protect \emph
  {positive-definite energy} in our Universe with $\Lambda >0$, within the
  \protect \emph {full non-linear} theory of general relativity \cite
  {Vee2018}.

\bibitem{Pen63}
R.~Penrose, Phys. Rev. Lett. \textbf{10}, 66 (1963).
\newblock \doi{10.1103/PhysRevLett.10.66}.
\newblock \urlprefix\url{http://link.aps.org/doi/10.1103/PhysRevLett.10.66}

\bibitem{Pen65}
R.~Penrose, Proceedings of the Royal Society of London A: Mathematical,
  Physical and Engineering Sciences \textbf{284}(1397), 159 (1965).
\newblock \doi{10.1098/rspa.1965.0058}

\bibitem{Pen88}
R.~Penrose, W.~Rindler, \emph{Spinors and Space-Time: Volume 2, Spinor and
  Twistor Methods in Space-Time Geometry} (Cambridge Monographs on Mathematical
  Physics, Cambridge, 1988)

\bibitem{Bondi60}
H.~Bondi, Nature \textbf{186}, 535 (1960)

\bibitem{GP}
J.~Griffiths, J.~Podolsk\'{y}, \emph{Exact Space-Times in Einstein's General
  Relativity} (Cambridge Monographs on Mathematical Physics, Cambridge, 2012)

\bibitem{Note2}
These two spin coefficients $\sigma '$ and $\kappa '$ necessarily vanish for
  the Schwarzschild-(anti-)de Sitter spacetime (which is spherically symmetric)
  since they have non-zero spin weights. Well, the static case only served as
  an investigation that suggested what a suitable set of fall-offs might be.
  The eventual validation for the correct set of fall-offs came from being able
  to successfully generalise the asymptotic solutions for $\Lambda =0$ \cite
  {newunti62,Pen88} to include a non-zero $\Lambda $ with the physics of
  gravitational waves \cite {Vee2016}. This incidentally, does not require
  polyhomogeneous terms.

\bibitem{Szabados}
L.B. Szabados, P.~Tod, Classical and Quantum Gravity \textbf{32}(20), 205011
  (2015).
\newblock \urlprefix\url{http://stacks.iop.org/0264-9381/32/i=20/a=205011}

\bibitem{Note3}
The $\protect \mathaccentV {vec}17E{l}$ and $\protect \mathaccentV {vec}17E{n}$
  null tetrad vectors in Ref. \cite {Szabados} are related to those in Refs.
  \cite {Vee2016,Vee2017} by a boost transformation. Apart from that, the
  $\protect \mathaccentV {vec}17E{m}$ and $\protect \mathaccentV
  {vec}17E{\protect \mathaccentV {bar}016{m}}$ null tetrad vectors in the
  former are prescribed to have no $\protect \mathaccentV {vec}17E{\partial
  }_r$ component, whilst those in the latter papers are instead required to be
  parallel transported along $\protect \mathaccentV {vec}17E{l}$ --- in the
  spirit of Refs. \cite {newpen62,newunti62}. These can be related via a null
  rotation around $\protect \mathaccentV {vec}17E{l}$. Therefore, some work is
  required to convert the fall-offs from one set of null tetrad to the other.
  Note also that a smooth conformal compactifiability would rule out
  polyhomogeneous terms --- and is adequate to describe gravitational
  radiation.

\bibitem{Note4}
We note that with a cosmological constant, the peeling property is not quite an
  invariant concept, since null infinity $\protect \mathcal {I}$ is not a null
  hypersurface for $\Lambda \not =0$ and so there is no well-defined null
  direction \cite {Pen88}. Nevertheless within the context of an isolated
  system with $\Lambda >0$, this isolated system defines a timelike infinity
  $i^+$ on the spacelike $\protect \mathcal {I}$ and has a null hypersurface
  defining its cosmological horizon. The peeling property is then well-defined,
  with respect to this isolated system and the radiation emitted by it. See
  Ref. \cite {Vee2017d} for a discussion.

\bibitem{200years}
A.~Ashtekar, L.~Bombelli, O.~Reula, in \emph{The Covariant Phase Space of
  Asymptotically Flat Gravitational Fields}, ed. by M.~Francaviglia, D.~Holm
  (North-Holland, Amsterdam, 1990), chap. Mechanics, Analysis and Geometry: 200
  Years after Lagrange

\bibitem{Mao19}
P.~Mao, Phys. Rev. D \textbf{99}, 104024 (2019).
\newblock \doi{10.1103/PhysRevD.99.104024}.
\newblock \urlprefix\url{https://link.aps.org/doi/10.1103/PhysRevD.99.104024}

\bibitem{Com19}
G.~Comp\`{e}re, A.~Fiorucci, R.~Ruzziconi, Classical and Quantum Gravity
  \textbf{36}(19), 195017 (2019).
\newblock
  \urlprefix\url{https://iopscience.iop.org/article/10.1088/1361-6382/ab3d4b}

\bibitem{Note5}
The different fall-offs appearing here, as compared to the asymptotically flat
  case, are $\rho '^o_{-1}$, $\gamma ^o_{-1}$, as well as the $O(1)$ order
  terms for $\sigma '$ and $\kappa '$. The first two are expected from the case
  of purely vacuum (anti-)de Sitter spacetime, where $\rho '^o_{-1}=-\Lambda
  /6$ and $\gamma ^o_{-1}=-\Lambda /6$. The requirement of an $O(1)$ term for
  $\sigma '$ was found from the spin coefficient equation involving $D\sigma
  '$, otherwise the leading order term for the shear $\sigma $ that represents
  outgoing gravitational waves would vanish. After a substantially long
  sequence of calculations, it was found that a strange condition on the
  leading order term for $\sigma $ came up, which went away if $\kappa '$ has
  an $O(1)$ term. Again, those new leading order terms due to $\Lambda $ are
  consistent with the results presented in Ref. \cite {Szabados}.

\bibitem{Note6}
Incidentally, a smooth conformal compactifiability and being vacuum (possibly
  with $\Lambda $) near $\protect \mathcal {I}$ would imply the vanishing of
  the Weyl curvature on $\protect \mathcal {I}$ \cite {Pen88}.

\bibitem{Note7}
In the derivation of the asymptotic solutions with $\Lambda $ as detailed in
  the appendices of Refs. \cite {Vee2016,Vee2017}, this Bianchi identity is the
  fourth last equation to be solved, i.e. the 35th out of the 38 Newman-Penrose
  equations (or the 39th out of the 42 Newman-Penrose-Maxwell equations). As
  the focus there was to produce the mass-loss formula instead of analysing the
  peeling property, it was not at all obvious that Eq. (\ref {Psi4}) would give
  $(\Psi _0)^o_4=0$ [there, the trivial equation ``$0=0$'' was obtained, since
  $\Psi _0=O(r^{-5})$ was assumed].

\bibitem{Pen87}
R.~Penrose, W.~Rindler, \emph{Spinors and Space-Time: Volume 1, Two-Spinor
  Calculus and Relativistic Fields} (Cambridge Monographs on Mathematical
  Physics, Cambridge, 1987)

\bibitem{Note8}
Well, the conditions $\protect \mathaccentV {bar}016{g}_{\protect \mathaccentV
  {bar}016{r}\protect \mathaccentV {bar}016{\theta }}$ and $\protect
  \mathaccentV {bar}016{g}_{\protect \mathaccentV {bar}016{u}\protect
  \mathaccentV {bar}016{r}}$ can be solved for $a^1$ and $g^1$. These may be
  plugged into $\protect \mathaccentV {bar}016{g}_{\protect \mathaccentV
  {bar}016{r}\protect \mathaccentV {bar}016{r}}$ to show that it is trivially
  satisfied.

\bibitem{Ortaggio1}
M.~Ortaggio, A.~Pravdov\'a, Phys. Rev. D \textbf{90}, 104011 (2014).
\newblock \doi{10.1103/PhysRevD.90.104011}.
\newblock \urlprefix\url{https://link.aps.org/doi/10.1103/PhysRevD.90.104011}

\bibitem{Ortaggio2}
M.~Ortaggio, Phys. Rev. D \textbf{90}, 124020 (2014).
\newblock \doi{10.1103/PhysRevD.90.124020}.
\newblock \urlprefix\url{https://link.aps.org/doi/10.1103/PhysRevD.90.124020}

\bibitem{gracos1}
N.T. Bishop, Phys. Rev. D \textbf{93}, 044025 (2016).
\newblock \doi{10.1103/PhysRevD.93.044025}.
\newblock \urlprefix\url{http://link.aps.org/doi/10.1103/PhysRevD.93.044025}

\bibitem{Zhang}
F.~Xie, X.~Zhang, ``Peeling property of Bondi-Sachs metrics for nonzero
  cosmological constant''  (2017).
\newblock \urlprefix\url{https://arxiv.org/abs/1704.06015}

\bibitem{Note9}
I thank Xiao Zhang for explicitly pointing this out to me, in terms of the
  Newman-Unti null tetrad that I have employed.

\bibitem{Rod}
A.~{\v C}ap, A.~Gover, Journal für die reine und angewandte Mathematik
  \textbf{717}, 47 (2016).
\newblock \doi{10.1515/crelle-2014-0036}.
\newblock
  \urlprefix\url{https://www.degruyter.com/view/j/crelle.2016.2016.issue-717/crelle-2014-0036/crelle-2014-0036.xml}

\bibitem{Note10}
Private communication with J{\"o}rg Frauendiener.

\bibitem{ash3}
A.~Ashtekar, B.~Bonga, A.~Kesavan, Phys. Rev. D \textbf{92}, 104032 (2015).
\newblock \doi{10.1103/PhysRevD.92.104032}.
\newblock \urlprefix\url{http://link.aps.org/doi/10.1103/PhysRevD.92.104032}

\bibitem{Note11}
Recall that our $r$ is an affine parameter of the null generators tangent to
  the null tetrad vector $\protect \mathaccentV {vec}17E{l}$, which is an
  outgoing null vector. Then, $r\rightarrow \infty $ goes to infinity along a
  null direction, i.e. approaches future null infinity $\protect \mathcal {I}$.

\bibitem{gracos2}
G.~Date, S.J. Hoque, Phys. Rev. D \textbf{94}, 064039 (2016).
\newblock \doi{10.1103/PhysRevD.94.064039}.
\newblock \urlprefix\url{http://link.aps.org/doi/10.1103/PhysRevD.94.064039}

\end{thebibliography}

\end{document}